\documentclass[onecollarge,natbib]{svjour2}
\bibpunct{[}{]}{;}{n}{}{,} 
\smartqed  
\usepackage{graphicx}
\usepackage{url}
\usepackage{amsfonts,amssymb}
\usepackage{bm}
\journalname{}
\begin{document}

\title{Light-quark baryon spectroscopy within ANL-Osaka dynamical coupled-channels approach\thanks{
This work was supported by the JSPS KAKENHI Grant No. 25800149,
and by the HPCI Strategic Program (Field 5 ``The Origin of Matter and the Universe'')
of MEXT of Japan.}
}
\author{Hiroyuki Kamano}
\institute{H. Kamano \at
              Research Center for Nuclear Physics (RCNP), Osaka University, Ibaraki, Osaka 567-0047, Japan \\
              \email{kamano@rcnp.osaka-u.ac.jp}            \\
}

\date{Received: date / Accepted: date}

\maketitle

\begin{abstract}
Recent results on the study of light-quark baryons with the ANL-Osaka dynamical coupled-channels (DCC) approach are presented,
which contain the $N^*$ and $\Delta^*$ spectroscopy via the analysis of $\pi N$ and $\gamma N$ reactions and
the $\Lambda^*$ and $\Sigma^*$ spectroscopy via the analysis of $K^- p$ reactions.
A recent application of our DCC approach to neutrino-nucleon reactions in the resonance region is also presented.
\keywords{Excited baryons \and Dynamical coupled-channels model \and Partial-wave analysis \and Meson production reactions \and Neutrino nucleon reactions}
\end{abstract}

\section{Introduction}
\label{intro}

The spectroscopic study of light-quark baryons ($N^*$, $\Delta^*$, $\Lambda^*$, $\Sigma^*$) 
remains a central issue in the hadron physics.
Since the excitation spectra of light-quark baryons are an embodiment of
the dynamics of confinement in Quantum Chromodynamics (QCD),
their precise determination through analyzing reaction data is essential to
test QCD in the nonperturbative domain.
In this regard, several attempts to link the (real) energy spectrum of QCD in the finite box to 
the complex pole masses of actual hadron resonances have been being made recently
(see, e.g., Refs.~\cite{dudek,wu,molina,lee}).

A particular complexity exists in extracting the light-quark baryons from reaction data.
This originates from a peculiar character of the light-quark baryons:
they are broad and highly overlapping resonances.
The light-quark baryons are thus strongly correlated with 
each other and with backgrounds through reaction processes 
over the wide energy range,
and the observed shape of cross sections and spin asymmetries 
represents such a complicated interference between them.
As a result, the light-quark baryon resonances do not produce any clean isolated peak 
in cross sections with a few exceptions.
This makes identification of light-quark baryons from experiments rather difficult,
and cooperative works between experiments and theoretical analyses are indispensable
for the light-quark baryon spectroscopy.

To disentangle complicated correlations between the light-quark baryon resonances, 
one needs to perform a partial-wave analysis of various meson-production reactions 
comprehensively over the wide energy range.
Here the use of a reaction framework satisfying multichannel unitarity 
is particularly important.
In fact, as is well known, multichannel unitarity ensures the conservation of 
probabilities in the multichannel reaction processes, and this is crucial for accomplishing 
comprehensive analysis of meson-production reactions with various final states consistently within
a single reaction framework.
Also, the multichannel unitarity properly defines 
the analytic structure (branch points, cuts, threshold cusps, etc.) 
of scattering amplitudes in the complex-energy plane, as required by the scattering theory, which
is essential to extract resonance information from reaction data {\it correctly}.

As an approach that enables such a multichannel analysis,
a dynamical coupled-channels (DCC) model for meson-production reactions in the resonance region
was developed in Ref.~\cite{msl}.
So far, this model has been employed in a series 
of works~\cite{jlms07,jlmss07,djlss08,kjlms09,jklmss09,kjlms09-2,sjklms10,knls10,knls13,kamano13}
to analyze the $\pi N$, $\gamma N$, and $e N$ reactions
and study the mass spectrum, structure, and dynamical origins of 
nonstrange $N^*$ and $\Delta^*$ resonances.
This model has also been applied to the study of
$\Lambda^*$ and $\Sigma^*$ resonances with strangeness $S=-1$
through a comprehensive analysis of $K^- p$ reactions~\cite{knls14,knls15}.
In addition to these studies related to the light-quark baryon spectroscopy,
we have recently made an application of our DCC model to
neutrino-nucleon reactions in the resonance region~\cite{knls12,nks15}.
The purpose of this contribution is to give an overview of our recent efforts for 
these subjects.

\section{ANL-Osaka DCC model}
\label{sec:1}

The basic formula of our DCC model is the coupled-channels integral equation 
obeyed by the partial-wave amplitudes of $a \to b$ reactions
that are specified by the total angular momentum ($J$), parity ($P$), and total isospin ($I$)
(here we explain our approach by taking the $N^*$ and $\Delta^*$ sector as an example):
\begin{equation}
T^{(J^P I)}_{b,a} (p_b,p_a;W) = 
V^{(J^PI)}_{b,a} (p_b,p_a;W)
+\sum_c \int_C dp_c\,p_c^2 V^{(J^PI)}_{b,c} (p_b,p_c;W) G_c(p_c;W) T^{(J^PI)}_{c,a} (p_c,p_a;W),
\label{lseq}
\end{equation}
with
\begin{eqnarray}
V^{(J^P I)}_{b,a}(p_b,p_a; W)= 
v_{b,a}(p_b,p_a)+
Z_{b,a}(p_b,p_a;W)+
\sum_{N^*_n}\frac{\Gamma_{b,N^*_n}(p_b)
 \Gamma_{N^*_n,a}(p_a)} {W-M^0_{N^*_n}} .
\label{eq:veq}
\end{eqnarray}
Here the subscripts ($a,b,c$) represent the considered reaction channels (the indices associated
with the total spin and orbital angular momentum of the channels are suppressed);
$p_a$ is the magnitude of the relative momentum for the channel $a$ in the center-of-mass frame; and
$W$ is the total scattering energy.
For the $N^*$ and $\Delta^*$ sector, we have taken into account
the eight channels, $\gamma^{(*)}N$, $\pi N$, $\eta N$, $K\Lambda$, 
$K\Sigma$, $\pi\Delta$, $\rho N$, and $\sigma N$\footnote{
Because of the perturbative nature of the electromagnetic interactions,
it is only necessary to solve the coupled-channels equations
in the channel space excluding the $\gamma^{(*)} N$ channel. 
Thus the summation in Eq.~(\ref{lseq}) runs over only the hadronic channels.}, 
where the last three are the quasi two-body channels 
that subsequently decay into the three-body $\pi\pi N$ channel.
The Green's functions for the meson($M$)-baryon($B$) channels are given by
$G_c(p_c;W) = 1/[W-E_M(p_c)-E_B(p_c)+i\varepsilon]$ for 
$c=\pi N, \eta N, K\Lambda, K\Sigma$, while
$G_c(p_c;W) = 1/[W-E_M(p_c)-E_B(p_c)+\Sigma_{MB}(p_c;W)]$ for 
$c=\pi \Delta, \sigma N, \rho N$, where
$E_{\alpha}(p) = \sqrt{m_\alpha^2+p^2}$ is the relativistic single-particle energy 
of a particle $\alpha$,
and $\Sigma_{MB}(p_c;W)$ is the self-energy 
that produces the three-body $\pi \pi N$ cut.
The transition potential $V_{b,a}$ [Eq.~(\ref{eq:veq})] consists of three pieces.
The first two, $v_{b,a}$ and $Z_{b,a}$, describe the nonresonant processes including only 
the ground-state mesons and baryons belonging to each flavor SU(3) multiplet, 
and the third one describes the propagation 
of the bare $N^*$ states.
In our approach, the unitary transformation method~\cite{msl,sl}
is used to derive the potential $v_{b,a}$ from effective Lagrangians.
With this method, the resulting $v_{b,a}$ becomes energy independent and
its off-shell behavior is specified.
On the other hand, the $Z$-diagram potential, $Z_{b,a}$, is derived using
the projection operator method~\cite{feshbach}. 
It also produces the three-body $\pi \pi N$ cut~\cite{msl,knls13}, 
and implementation of both the $Z$-diagram potential 
and the self-energy in the Green's functions 
is necessary to maintain the three-body unitarity.
Furthermore, off-shell rescattering effects are also taken 
into account properly through the momentum integral in Eq.~(\ref{lseq}),
which are usually neglected in on-shell approaches.

\section{Recent efforts for $N^*$ and $\Delta^*$ spectroscopy with ANL-Osaka DCC model}
\label{sec:nucleon}

Our latest published model for the $N^*$ and $\Delta^*$ sector~\cite{knls13,kamano13} was 
constructed by making a simultaneous analysis of 
$\pi p \to \pi N, \eta N, K \Lambda, K \Sigma$ 
and $\gamma p \to \pi N, \eta N, K \Lambda, K \Sigma$.
This contains the data of both unpolarized differential cross sections and polarization 
observables up to $W = 2.1$ GeV (up to $W= 2.3$ GeV for $\pi p \to \pi N$),
which results in fitting $\sim 23,000$ data points.
After the completion of this analysis, 
we have been mainly proceeding with two subjects for the $N^*$ and $\Delta^*$ spectroscopy:
(1) extraction of the helicity amplitudes for the $\gamma n \to N^*$ transition
from the available $\gamma \textrm{`}n\textrm{'} \to \pi N$ data, and 
(2) determination of $Q^2$ dependence of the $p$-$N^*$ and $p$-$\Delta^*$ electromagnetic 
transition form factors by analyzing the structure function data for single-pion electroproductions off the proton target.
The latter subject is discussed in Ref.~\cite{tsato},
and in this contribution we focus on presenting the current status of 
our $\gamma \textrm{`}n\textrm{'} \to \pi N$ analysis.

There are mainly two reasons for studying meson photo- and electro-production reactions 
off the neutron target.
One comes from the fact that the analysis of the data for both proton- and neutron-target 
reactions is required to decompose the matrix elements for the electromagnetic currents 
into the ones of isoscalar and isovector currents and uniquely determine the isospin structure of 
the $\gamma^{(*)} N \to N^*$ transition amplitudes.
Another is because the matrix elements for such ``isospin-decomposed'' currents
are necessary for constructing a model for neutrino-induced reactions, which
will be discussed in Sec.~\ref{sec:neutrino}.

\begin{figure}[t]
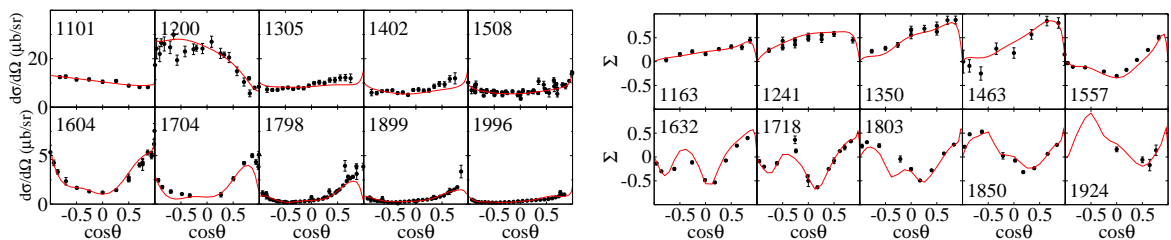

\centering
\includegraphics[clip,width=0.48\textwidth]{gnpi-p_dc.eps}
\ \
\includegraphics[clip,width=0.48\textwidth]{gnpi-p_s.eps}
\caption{\label{gn-obs1}
Differential cross sections (left) and photon asymmetries (right)
for 
$\gamma \textrm{`}n\textrm{'} \to \pi^- p$.
The numbers shown in each panel are the corresponding total scattering energy $W$ in MeV.
The data are taken from Ref.~\cite{gwu}.
}
\end{figure}

\begin{figure}[t]
\centering
\includegraphics[clip,width=0.48\textwidth]{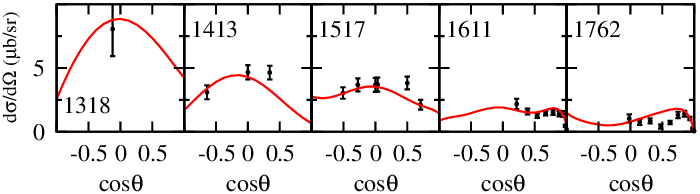}
\ \
\includegraphics[clip,width=0.48\textwidth]{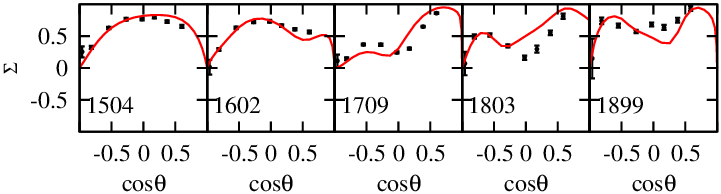}
\caption{\label{gn-obs2}
Differential cross sections (left) and photon asymmetries (right)
for 
$\gamma \textrm{`}n\textrm{'} \to \pi^0 n$.
The numbers shown in each panel are the corresponding total scattering energy $W$ in MeV.
The data are taken from Ref.~\cite{gwu}.
}
\end{figure}

So far, we have performed the fits to the data for $\gamma n \to \pi N$ up to $W\lesssim 2$ GeV. 
The data are available for $d\sigma/d\Omega$, $\Sigma$, $T$, and $P$ for $\gamma n \to \pi^- p$,
and $d\sigma/d\Omega$ and $\Sigma$ for $\gamma n \to \pi^0 n$.
This contains $\sim 3,200$ data points.
Some of the results of our fits are presented in Figs.~\ref{gn-obs1} and~\ref{gn-obs2}.
One can see that a reasonably good reproduction of the $\gamma n \to \pi N$ data
has been accomplished for the considered energy region.

\begin{figure}[t]
\centering
\includegraphics[clip,width=0.65\textwidth]{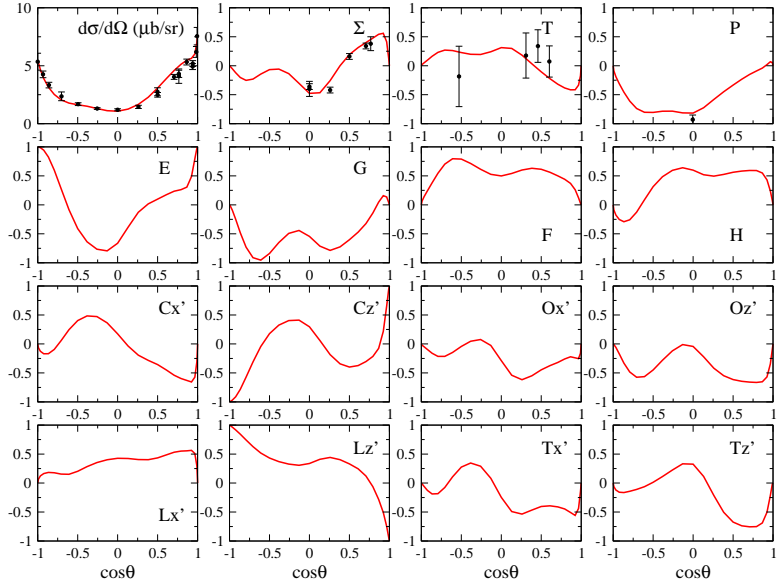}
\caption{\label{allobs}
Differential cross section and all polarization observables 
for $\gamma n \to \pi^- p$ at $W = 1604$ MeV ($E_\gamma =900$ MeV).
The results are predictions from our current DCC model except for $d\sigma/d\Omega$, $\Sigma$, $T$, and $P$.
The data are taken from Ref.~\cite{gwu}.
}
\end{figure}

In Fig.~\ref{allobs}, we present our calculated results for all 16 observables 
($d\sigma/d\Omega$ and 15 polarization observables) of
the $\gamma n \to \pi^- p$ reaction at $W = 1604$ MeV.
They are predictions from our current DCC model except for
$d\sigma/d\Omega$, $\Sigma$, $T$, and $P$, whose data displayed in Fig.~\ref{allobs}
were included in our fits to determine the model parameters.
The polarization data are known to provide crucial constraints on 
multipole amplitudes~\cite{shkl11}, and
tremendous efforts are now being pursued
at electron- and photon-beam facilities such as JLab, ELSA, and MAMI
to extract polarization observables for the neutron-target 
photoproductions through the $\gamma d$ reaction (see, e.g., Ref.~\cite{exp}).

\section{Comprehensive analysis of $K^- p$ reactions and extraction of $\Lambda^*$ and $\Sigma^*$ resonances}
\label{sec:hyperon}

Following the success of our $N^*$ and $\Delta^*$ studies,
we have recently extended our DCC approach to the strangeness $S=-1$ sector 
to also explore the $\Lambda^*$ and $\Sigma^*$ hyperon resonances.
In the formulation of the DCC model in the $S=-1$ sector, we have taken into account
seven channels ($\bar K N$, $\pi \Sigma$, $\pi \Lambda$, $\eta \Lambda$, 
$K\Xi$, $\pi \Sigma^*$, $\bar K^* N$), where
the $\pi \Sigma^*$ and $\bar K^* N$ channels are the quasi two-body channels 
that subsequently decay into the three-body $\pi\pi\Lambda$ and $\pi \bar K N$ channels, 
respectively.
The model parameters are then fixed by fitting to all available data of 
$K^- p \to \bar K N, \pi \Sigma, \pi \Lambda, \eta \Lambda, K\Xi$ 
reactions from the threshold up to $W = 2.1$ GeV.
The data contain the total cross section ($\sigma$), 
differential cross section ($d\sigma/d\Omega$), and recoil polarization ($P$), 
and this results in fitting more than 17,000 data points.

\begin{figure}[t]
\centering
\includegraphics[clip,width=0.7\textwidth]{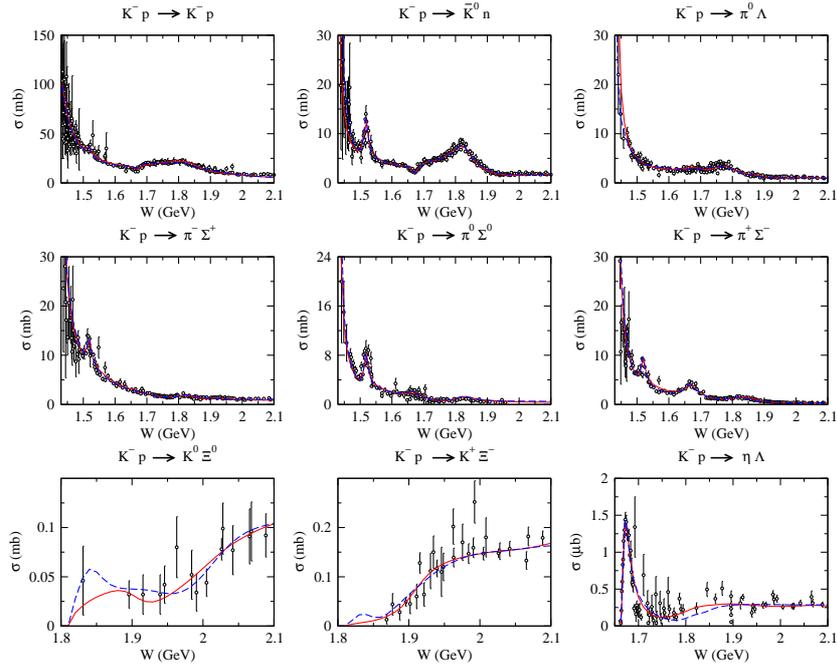}
\caption{
Fitted results for the total cross sections of $K^- p$ reactions~\cite{knls14}.
Red solid (blue dashed) curves are the results of Model A (Model B).
}
\label{fig:kptcs}
\end{figure}

Figure~\ref{fig:kptcs} shows the results of our fits to the total cross sections for all reactions we considered.
Here two curves are plotted in red (Model A) and blue (Model B), respectively.
This is because the existing $K^-p$ reaction data are not sufficient 
to determine our model parameters unambiguously, and it allowed us to have 
two distinct sets of the model parameters, yet both give almost the same $\chi^2$ value.
We quote Ref.~\cite{knls14} for all the details of our analysis including 
the fits to the differential cross sections and polarization observables.

Once the model parameters are determined by fitting to the reaction data,
we can extract various information (mass, width, pole residues, 
and branching ratios, etc.) on the $\Lambda^*$ and $\Sigma^*$ resonances
by making an analytic continuation of the calculated scattering amplitudes 
in the complex energy plane
(see Refs.~\cite{ssl,ssl2} for a numerical procedure to perform the analytic continuation within our DCC model).
This was done in Ref.~\cite{knls15}, and
the value of the resonance pole masses extracted obtained via such an analytic continuation 
is presented in Table~\ref{tab:resmass}.
We searched pole positions of scattering amplitudes in the complex energy region with
$m_{\bar K} + m_N \leq {\rm Re}(W) \leq 2.1$ GeV and
$0 \leq -{\rm Re}(W) \leq 0.2$ GeV, and found
18 resonances (10 $\Lambda^*$ and 8 $\Sigma^*$) in Model A, while
20 resonances (10 $\Lambda^*$ and 10 $\Sigma^*$) in Model B.
The uncertainties assigned for the value of each extracted pole mass was deduced by following a procedure similar to 
the one used in Ref.~\cite{shkl11} to determine the uncertainties in extracted multipole amplitudes 
for $\gamma p \to K^+ \Lambda$.
Here we did not assign the uncertainty for $J^P=7/2^+$ $\Lambda$ resonance in Model A because
it was turned out to be too large to be meaningful.

\begin{table}[t]
\caption{\label{tab:resmass}
Extracted complex pole masses ($M_R$) for the $\Lambda^*$ and $\Sigma^*$ resonances
found in the energy region above the $\bar K N$ threshold~\cite{knls15}.
The masses are listed as $\bm{(} {\rm Re}(M_R), -{\rm Im}(M_R)\bm{)}$
together with their deduced uncertainties.
The resonance poles are searched in the complex $W$ region with 
$m_{\bar K} + m_N \leq {\rm Re}(W) \leq 2.1$ GeV and $0\leq -{\rm Im}(W) \leq 0.2$ GeV,
and all of the resonances listed are located in the complex $W$ Riemann surface
nearest to the physical real $W$ axis. 
}
\centering
\begin{tabular}{cccccc}
\hline\noalign{\smallskip}
              & \multicolumn{2}{c}{$\Lambda^*$}  &
              & \multicolumn{2}{c}{$\Sigma^*$}  \\[3pt]
              & \multicolumn{2}{c}{$M_R$ (MeV)}  &
              & \multicolumn{2}{c}{$M_R$ (MeV)}  \\
\cline{2-3} \cline{5-6}
 $J^P(l_{I2J})$ & Model A & Model B &
 $J^P(l_{I2J})$ & Model A & Model B \\
\tableheadseprule\noalign{\smallskip}
$1/2^-(S_{01})$&--                                    &(1512$^{+1}_{-1}$,185$^{+1}_{-2}$)    &$1/2^-(S_{11})$&--                                     &(1551$^{+2}_{-9}$,188$^{+6}_{-1}$)\\ 
               &(1669$^{+3}_{-8}$, 9$^{+9}_{-1}$)     &(1667$^{+1}_{-2}$,12$^{+3}_{-1}$)     &               &(1704$^{+3}_{-6}$, 43$^{+7}_{-2}$)     &--\\
$1/2^+(P_{01})$&(1544$^{+3}_{-3}$, 56$^{+6}_{-1}$)    &(1548$^{+5}_{-6}$, 82$^{+7}_{-7}$)    &               &    --                                 &(1940$^{+2}_{-2}$, 86$^{+2}_{-2}$)\\
               &--                                    &(1841$^{+3}_{-4}$, 31$^{+3}_{-2}$)    &$1/2^+(P_{11})$&--                                      &(1457$^{+5}_{-1}$, 39$^{+1}_{-4}$)\\
               &(2097$^{+40}_{-1}$, 83$^{+32}_{-6}$)  &--                                    &               &(1547$^{+111}_{-59}$, 92$^{+43}_{-39}$)&--\\
$3/2^+(P_{03})$& --                                   &(1671$^{+2}_{-8}$, 5$^{+11}_{-2}$)    &               &--                                     &(1605$^{+2}_{-4}$, 96$^{+1}_{-5}$)\\
               &(1859$^{+5}_{-7}$, 56$^{+10}_{-2}$)   & --                                   &               &(1706$^{+67}_{-60}$, 51$^{+79}_{-42}$) &--\\
$3/2^-(D_{03})$&(1517$^{+4}_{-4}$,  8$^{+5}_{-4}$)    &(1517$^{+4}_{-3}$, 8$^{+6}_{-6}$)     &               & --                                    &(2014$^{+6}_{-13}$, 70$^{+14}_{-1}$)\\
               &(1697$^{+6}_{-6}$, 33$^{+7}_{-7}$)    &(1697$^{+6}_{-5}$, 37$^{+7}_{-7}$)    &$3/2^-(D_{13})$&--                                     &(1492$^{+4}_{-7}$, 69$^{+4}_{-7}$)\\
$5/2^-(D_{05})$&(1766$^{+37}_{-34}$,106$^{+47}_{-31}$)& --                                   &               &(1607$^{+13}_{-11}$,126$^{+15}_{-9}$)  &--\\
               &(1899$^{+35}_{-37}$, 40$^{+50}_{-17}$)&(1924$^{+52}_{-24}$, 45$^{+57}_{-17}$)&               &(1669$^{+7}_{-7}$, 32$^{+5}_{-7}$)     &(1672$^{+5}_{-10}$, 33$^{+3}_{-3}$)\\
$5/2^+(F_{05})$&(1824$^{+2}_{-1}$, 39$^{+1}_{-1}$)    &(1821$^{+1}_{-1}$, 32$^{+1}_{-1}$)    &$5/2^-(D_{15})$&(1767$^{+2}_{-2}$, 64$^{+2}_{-1}$)     &(1765$^{+2}_{-1}$, 64$^{+3}_{-1}$)\\
$7/2^+(F_{07})$&(1757, 73)                            &--                                    &$5/2^+(F_{15})$&--                                     &(1695$^{+20}_{-77}$, 97$^{+50}_{-44}$)\\
               &--                                    &(2041$^{+80}_{-82}$,119$^{+57}_{-17}$)&               &(1890$^{+3}_{-2}$, 49$^{+2}_{-3}$)     &--\\
               &                                      &                                      &$7/2^+(F_{17})$&(2025$^{+10}_{-5}$, 65$^{+3}_{-12}$)   &(2014$^{+12}_{-1}$,103$^{+3}_{-9}$)\\
\noalign{\smallskip}\hline
\end{tabular}
\end{table}
\begin{figure}[t]
\centering
\includegraphics[clip,width=0.6\textwidth]{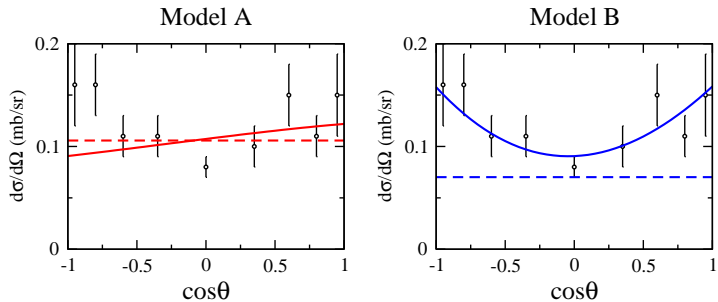}
\caption{
Differential cross section at $W=1672$ MeV for the $K^- p \to \eta \Lambda$ 
reaction.
Left (right) panel is the result from Model A (Model B).
Solid curves are the full results, while the dashed curves are the contribution
from the $S_{01}$ partial wave only.
For Model B, the difference between the solid and dashed curves almost comes from
the $P_{03}$ partial wave dominated by the new narrow 
$J^P=3/2^+$ $\Lambda$ resonance with $M_R = 1671^{+2}_{-8}-i(5^{+11}_{-2})$ MeV.
}
\label{fig:etldcs}
\end{figure}

We see in Table~\ref{tab:resmass} that the pole mass of 9 resonances (6 $\Lambda^*$ and 3 $\Sigma^*$)
agree well between our two models within the deduced uncertainties (each resonance is aligned in the same row).
These resonances would correspond to $\Lambda(1670)1/2^-$,
$\Lambda(1600)1/2^+$, $\Lambda(1520)3/2^-$, $\Lambda(1690)3/2^-$, $\Lambda(1820)5/2^+$, $\Lambda(1830)5/2^-$
$\Sigma(1670)3/2^-$, $\Sigma(1775)5/2^-$, and $\Sigma(2030)7/2^+$ in the PDG notation~\cite{pdg2014}.
They are assigned as four-star resonances by PDG except for $\Lambda(1600)1/2^+$ with three-star status.
However, within the energy region we are currently interested in, 
two more resonances, $\Lambda(1890)3/2^+$ and $\Sigma(1915)5/2^+$, are also rated as four-star by PDG,
but the corresponding resonances are not found in Model B.
This is one example indicating that four-star resonances rated by PDG
using the Breit-Wigner parameters
are not confirmed by the multichannel analyses in
which the resonance parameters are extracted at pole positions~\cite{knls15}.

A noteworthy difference between our two models is the existence of 
a new narrow $J^P=3/2^+$ $\Lambda^*$ resonance with $M_R = 1671^{+2}_{-8}-i (5^{+11}_{-2})$ MeV, 
which is found only in Model B.
We refer to this resonance as $\Lambda(1671)3/2^+$.
It is located close to the $\eta \Lambda$ threshold and has almost the same $M_R$ value as 
the four-star $\Lambda(1670)1/2^-$ resonance 
[see the second $J^P(l_{I2J}) = 1/2^- (S_{01})$ resonance in Table~\ref{tab:resmass}].
It is known that the sharp peak of the $K^- p \to \eta \Lambda$ total cross section near the threshold 
(see the bottom-right panel of Fig.~\ref{fig:kptcs})
can be explained by the existence of $\Lambda(1670)1/2^-$~\cite{bnl}.
In addition to this, we further find that the new $\Lambda(1671)3/2^+$ resonance seems to be 
favored by 
the differential cross section data (Fig.~\ref{fig:etldcs}).
Actually, both Models A and B reproduce the $K^- p \to \eta \Lambda$ total cross section equally well, but
Model A with no $\Lambda(1671)3/2^+$ does not reproduce the concave-up behavior of the 
differential cross section data well.

\section{Application to neutrino-induced reactions in resonance region}
\label{sec:neutrino}

\begin{figure}[t]
\centering
\includegraphics[clip,width=0.4\textwidth]{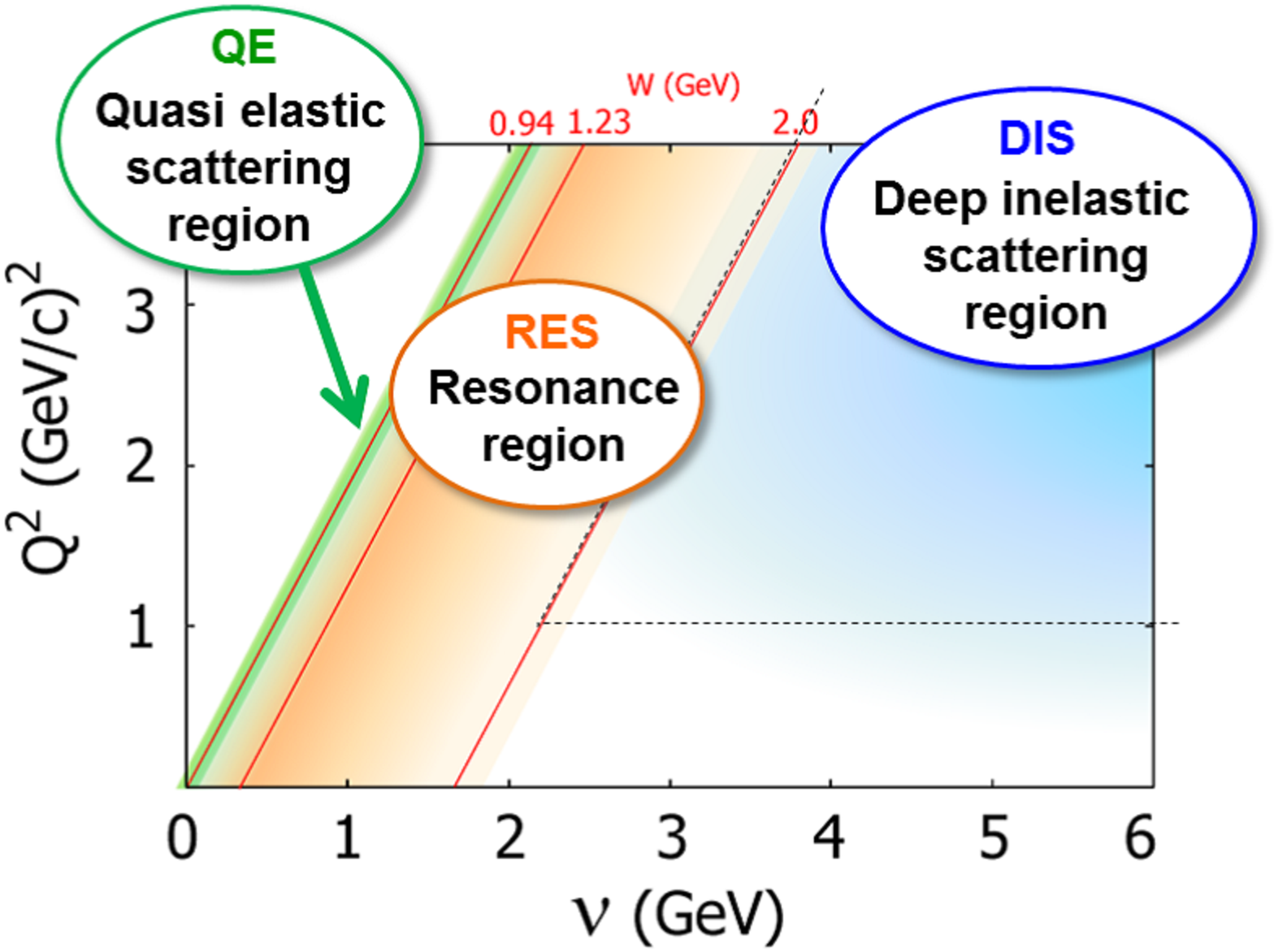}
\qquad  
\includegraphics[clip,width=0.55\textwidth]{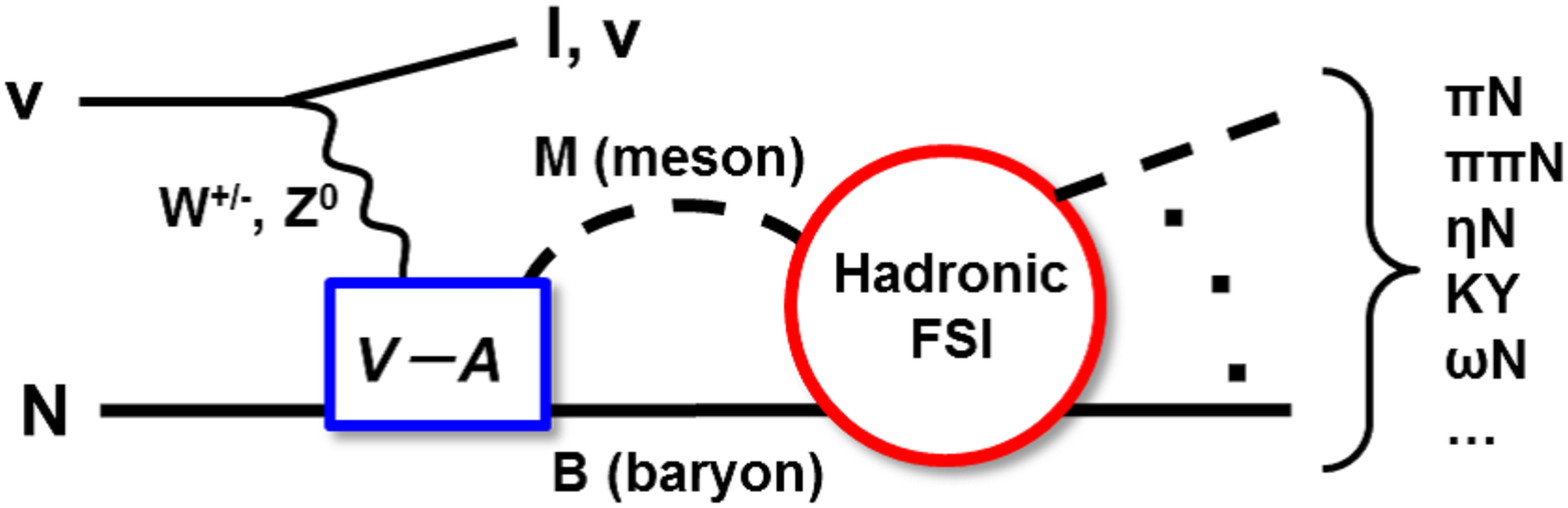}
\caption{
(Left) Kinematical region relevant to neutrino-oscillation experiments.
(Right) Schematic view of neutrino-nucleon reaction processes in resonance region.
}
\label{fig:neutrino1}
\end{figure}
\begin{figure}[t]
\centering
\includegraphics[clip,width=0.3\textwidth]{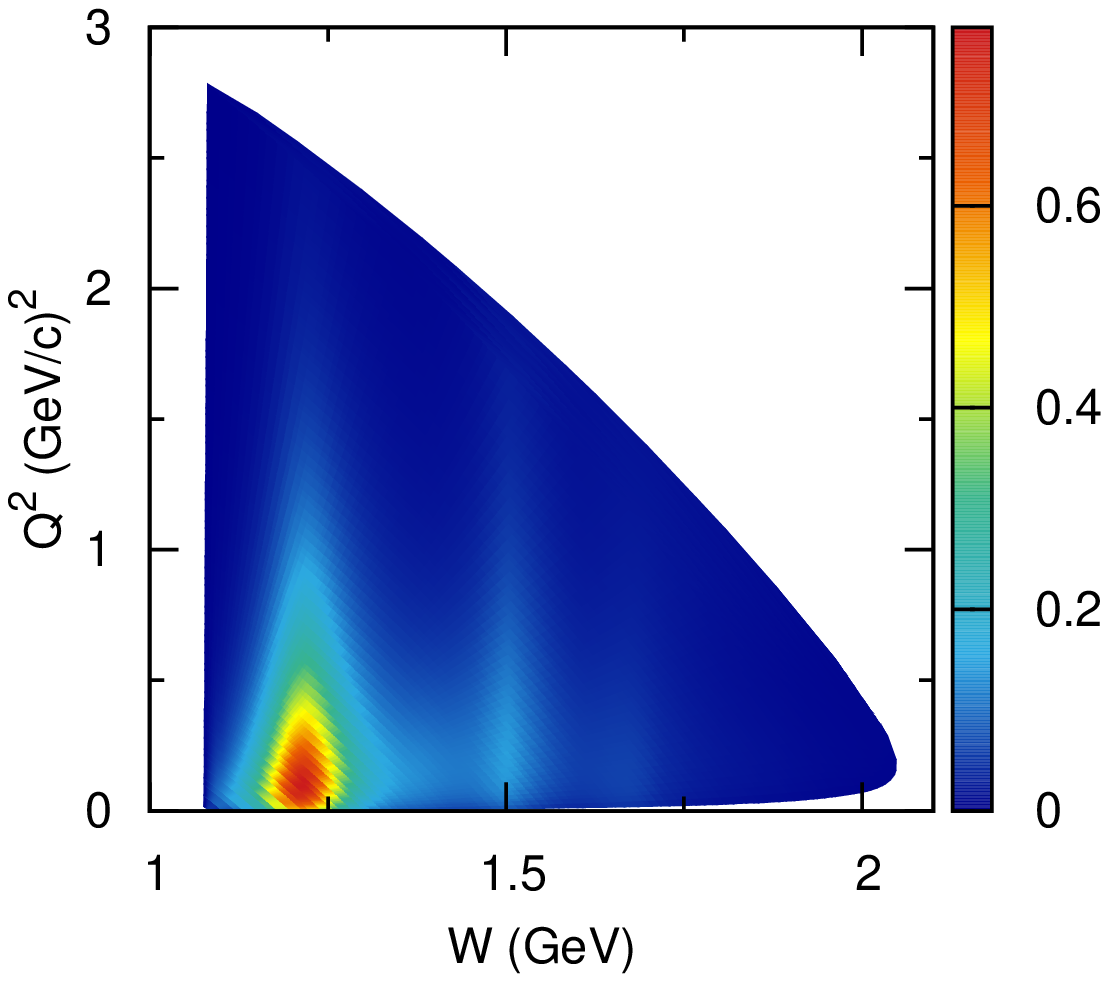}
\qquad  
\includegraphics[clip,width=0.3\textwidth]{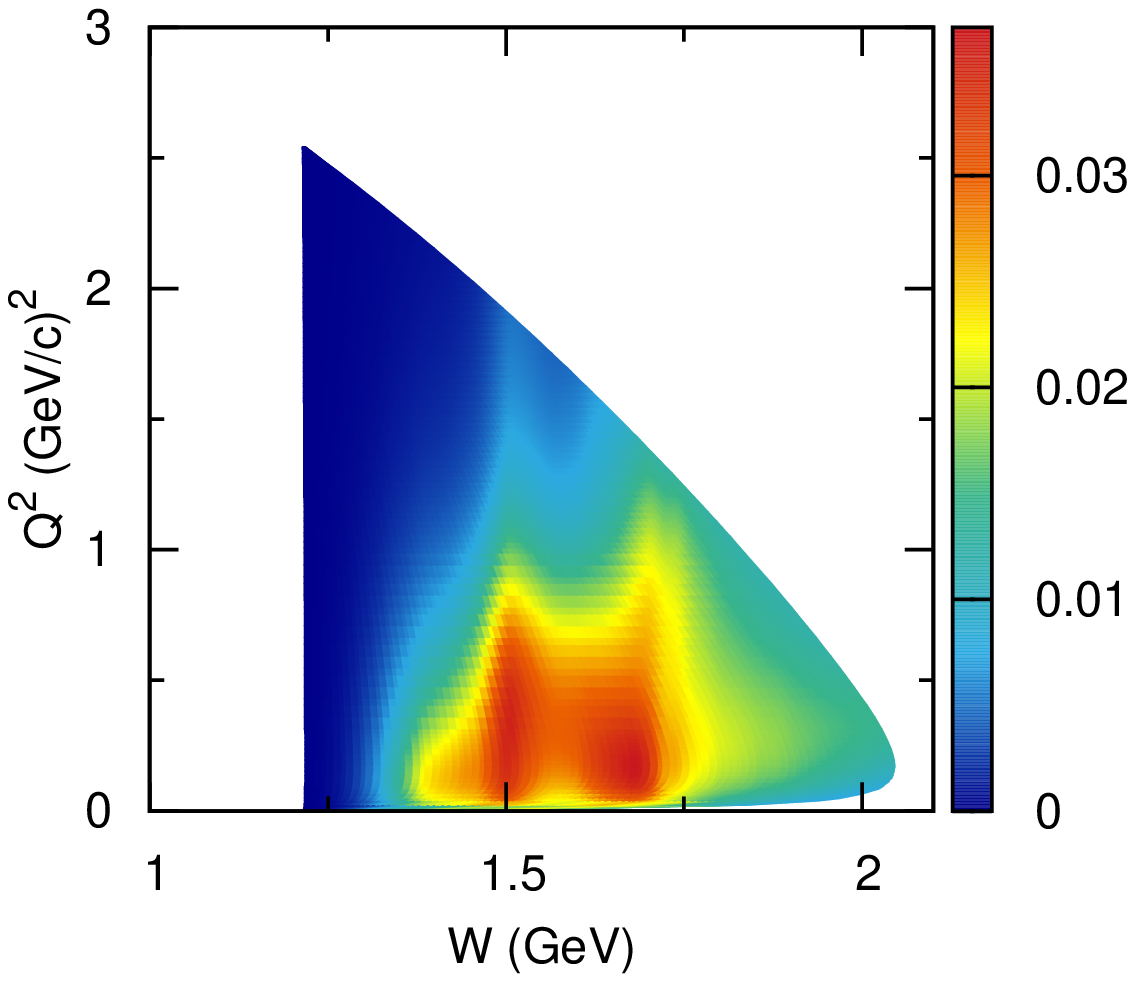}
\qquad  
\includegraphics[clip,width=0.25\textwidth]{neutrino-tcs.eps}
\caption{
(Left) $d\sigma/dWdQ^2$ for $\nu_\mu n \to \mu^- \pi N$ at $E_\nu = 2$ GeV.
(Middle) $d\sigma/dWdQ^2$ for $\nu_\mu n \to \mu^- \pi^+ \pi^- p$ at $E_\nu = 2$ GeV.
(Right) $\sigma$ for $\nu_\mu n \to \mu^- \pi N$ (black curve) and 
$\nu_\mu n \to \mu^- \pi \pi N$ (red curve) as a function of $E_\nu$.
See Ref.~\cite{nks15} for the details.
}
\label{fig:neutrino2}
\end{figure}

After the establishment of the neutrino mixing between all three flavors, 
major interests in the neutrino physics 
are shifting to the determination of the leptonic $CP$ phase and 
the neutrino-mass hierarchy (see, e.g., Ref.~\cite{hayato}).
In such neutrino-parameter searches via next-generation neutrino-oscillation experiments,
the lack of knowledge of neutrino-nucleus reaction cross sections
are expected to become one of the major sources of the systematic uncertainties.
Thus the developments of reaction models that give the cross sections 
in 10\% or better accuracy are now highly desirable.
Furthermore, the kinematics relevant to the neutrino-oscillation experiments 
extends over the quasi-elastic (QE), resonance (RES), 
and deep-inelastic-scattering (DIS) regions (see the left panel of Fig.~\ref{fig:neutrino1}), 
which are governed by rather different physics mechanisms and theoretical foundations.
These facts lead experimentalists and theorists in different fields to developing a new 
collaboration~\cite{kek,kek2} at the J-PARC Branch of KEK Theory Center, 
aiming at constructing a unified neutrino-reaction model that covers
all the relevant kinematical regions mentioned above consistently.
We will employ our DCC model as a base model for the RES region.

As a first step towards developing a model for neutrino-nucleus reactions in the RES region,
we have recently constructed a DCC model for neutrino-nucleon ($\nu N$) reactions~\cite{nks15}.
This is actually an important step because it can be used as an input to elementary 
processes in nuclear-target reactions.
The right panel of Fig.~\ref{fig:neutrino1} presents 
a schematic view of the $\nu N$ reactions in the RES region. 
The blue rectangle represents the matrix element for the transition between 
the nucleon and meson-baryon ($MB$) states induced by the (``$V-A$''-type) weak currents, 
while the red circle represents the rescattering process induced by 
the hadronic interactions.
In our DCC model, the rescattering part has been determined through 
the analysis of $\pi N$ and $\gamma N$ reaction data.
However, the weak-current matrix elements have to be newly constructed. 
The vector-current part can be determined using the data for photon- and electron-induced 
meson-production reactions off the proton and neutron targets 
(see Sec.~\ref{sec:nucleon} and Ref.~\cite{tsato}), and
in Ref.~\cite{nks15}, the vector-current matrix elements 
were determined in the range of $W \leq 2$ GeV and $Q^2 \leq 3$ GeV$^2$.
The axial-current matrix elements are difficult to be determined 
because plenty of $\nu N$ and/or $\nu d$ reaction data are required for that purpose, but
the existing data are far from sufficient to constrain our model parameters.
Therefore, at this stage, we determined the axial-current matrix elements 
by making use of PCAC hypothesis, and by imposing purely phenomenological 
assumptions for their $Q^2$ dependence, for which we will not go into the details here
and just quote Ref.~\cite{nks15}.
With this setup, we can make a prediction for the neutrino-nucleon reactions
in the RES region.

Figure~\ref{fig:neutrino2} shows the results for the charged-current (CC) single- and double-pion 
productions predicted from our DCC model.
This is the first-time fully coupled-channels calculation of $\nu N$
reactions beyond the $\Delta(1232)3/2^+$ region.
The contour plot of $d\sigma/dWdQ^2$ for $\nu_\mu n \to \mu^- \pi N$ 
(the left panel of Fig.~\ref{fig:neutrino2}) 
show a clear peak due to $\Delta(1232)3/2^+$ and relatively small peak at $W\sim 1.5$
that comes from $N(1535)1/2^-$ and $N(1520)3/2^-$ and so on.
In contrast, for the $\nu_\mu n \to \mu^- \pi^+ \pi^- p$ case,
the contribution from $\Delta(1232)3/2^+$ is barely seen, while higher resonance contributions
are sizable above $W\gtrsim 1.45$ GeV, and their effects on the cross section 
reach a rather high $Q^2$, too.
The single-pion productions are known to be a major background in the neutrino-oscillation
experiments at $E_\nu \lesssim 1$ GeV.
However, 
above $E_\nu \sim 1$ GeV, the cross section of the double-pion productions 
also becomes significant (the right panel of Fig.~\ref{fig:neutrino2}) 
because of the large branching ratios of higher $N^*$ and $\Delta^*$ resonances 
for the decay into the three-body $\pi \pi N$ states.
This implies the need of reliable information on the double-pion productions
for next-generation oscillation experiments using multi-GeV neutrino beams.

\section{Summary}
\label{sec:summary}

We have given an overview of our efforts for the light-quark baryon spectroscopy 
through a dynamical coupled-channels analysis of various meson-production reactions 
off the nucleon, induced by pion, photon, electron, and anti-kaon beams.
In particular, our recent DCC analysis of $K^- p$ reactions
has successfully extracted parameters associated with $\Lambda^*$ and $\Sigma^*$ resonances 
defined by poles of scattering amplitudes, and
has shed light on a possible existence of new 
$\Lambda^*$ and $\Sigma^*$ resonances.
However, our results also reveal that the current existing data for $K^- p$ reactions
are still far from sufficient to constrain the extracted partial-wave amplitudes and 
$\Lambda^*$ and $\Sigma^*$ resonance parameters unambiguously.
The help of hadron beam facilities such as J-PARC is highly desirable
for further establishing $\Lambda^*$ and $\Sigma^*$ resonances.
Finally, although we did not discuss in this contribution, 
our DCC approach has also been applied to meson spectroscopy~\cite{knls11,nkls12}.
We are putting more efforts into this direction, too.

\begin{acknowledgements}
The author thanks T.-S. H. Lee, S. X. Nakamura, and T. Sato for their collaborations.
\end{acknowledgements}

\end{document}